\documentclass[useAMS,usenatbib]{mn2e}
\usepackage{times,url,natbib,graphicx,amssymb}
\usepackage{subfigure}

\long\def\symbolfootnote[#1]#2{\begingroup%
\def\thefootnote{\fnsymbol{footnote}}\footnote[#1]{#2}\endgroup}

\begin{document}


\title
[IR Contamination in Galactic XRNe]
{IR Contamination in Galactic X-Ray Novae}
\author
[M.~Reynolds et al.]
{Mark T. Reynolds$^{1}$\thanks{email : m.reynolds@ucc.ie}, Paul
  J. Callanan$^{1}$, Edward L. Robinson$^{2}$ and Cynthia S. Froning$^{3}$ \\ 
$^{1}$Physics Department, University College Cork, Ireland\\
$^{2}$Department of Astronomy, University of Texas at Austin, Austin, TX
78712\\
$^{3}$Center for Astrophysics and Space Astronomy, University of
Colorado}

\maketitle


\begin{abstract}
The most widely used means of measuring the mass of black holes
in Galactic binaries - specifically the X-ray novae - involves both
radial velocity measurements of the secondary star, and photometric
measurements of its ellipsoidal variability. The latter is important
in constraining the inclination and mass ratio, and requires as direct
a measure of the flux of the secondary as possible. Up to now, such
measurements have been preferentially carried out in the NIR (1 --
2.5$\mu m$), where the flux from the cooler secondary is expected to 
dominate over that from the accretion disc. However, here we present
evidence of a significant non-stellar contribution to the NIR flux in
many of those quiescent X-ray novae that are thought to contain a black
hole primary. We discuss origins of this excess and the effect of
such contamination on Galactic black hole mass measurements.  
\end{abstract}



\section{Introduction}
X-ray novae (XRNe) are a subset of the low mass X-ray binaries
(LMXBs), containing a compact object primary (M$_1$) accreting material via
Roche lobe overflow from a low mass (M$_2 \lesssim$ 1 M$_{\sun}$)
secondary star. The secondary is normally a K or M-type main sequence
star, whereas the primary is either a neutron star or a black hole
with a characteristic mass of 1.4 M$_{\sun}$ or $\sim$ 10 M$_{\sun}$
respectively. These systems undergo quasi-periodic outbursts
(recurrence timescales $\sim$ tens of years) during which their
broadband luminosities (X-ray - radio) increase by many orders of
magnitude. They have proven to be a fertile hunting ground for the
discovery of black holes, with $\sim$ 75\% of the known stellar mass
black holes residing in these XRNe. See the reviews by \citet{b2},
\citet{b74} and \citet{b106} for further information about the X-ray,
UV/optical/IR and radio properties of these systems respectively.   

\begin{table*}
\caption{General properties of the various XRNe examined in this
  paper. See the papers indicated in column 1 for the general properties
  of each system. The veiling estimates are from the papers indicated in
  final column (see also \citealt{b74}).}
\label{extinct-table}
\begin{center}
\begin{tabular}{lcccccc}
\hline
Source & Spectral type & P$\rm_{orb}$ & Distance & Extinction & M$\rm_x$ &
R-band veiling \\
& & [ hrs] & [ Kpc ] &  E(B-V) & [ M$_{\sun}$ ] & \\ 
\hline\hline 
XTE J1118+480 $^{a}$ & K5V - M0V & 4.1  & 1.8 $\pm$ 0.6 &  0.02 $\pm$
0.006 & 8.53 $\pm$ 0.6 & 0.45 $^{h}$\\
GRO J0422+32 $^{b}$  & M0V - M2V & 5.1  & 2.7 $\pm$ 0.3 &  0.3 $\pm$ 0.1
& $\sim$ 10 & 0.39 $^{i}$\\
A0620-00 $^{c}$      & K3V - K7V & 7.8  & 1.2 $\pm$ 0.4 &  0.39 $\pm$
0.02 & 9.7 $\pm$ 0.6 & 0.10 $^{j, k}$\\ 
GS 2000+25 $^{d}$    & K3V - K6V & 8.3  & 2.7 $\pm$ 0.7 &  1.5 $\pm$ 0.1
& 5.5 - 8.8 & 0.32 \\
GS 1124-683 $^{e}$   & K3V - K5V & 10.4 & 5.5 $\pm$ 1.0 &  0.30 $\pm$
0.05 & 6.95 $\pm$ 0.6 & 0.15 $^{l}$\\
Cen X-4 $^{f}$       & K3V - K7V & 15.1 & 0.9 - 1.7 &  0.1 $\pm$ 0.05 & 1.5
$\pm$ 1.0 & 0.25 $^{m}$\\
GS 2023+338 $^{g}$      & K0IV      & 156  &  4$^{+2.0} _{-1.2}$ & 1.0
$\pm$ 0.1 & 12$^{+3} _{-2}$ & 0.16 $^{n}$\\
\hline
\end{tabular}
\end{center}
\medskip
{
REFERENCES: $^{a}$ \citealt{b5}; $^{b}$ \citealt{b21}; $^{c}$ \citealt{b20};
$^{d}$ \citealt{b47}; $^{e}$ \citealt{b13}; $^{f}$ \citealt{b65}; $^{g}$
\citealt{b58}; $^{h}$ \citealt{b41}; $^{i}$ \citealt{b80}; $^{j}$
\citealt{b77}; $^{k}$ \citealt{b86}; $^{l}$ \citealt{b78}; $^{m}$
\citealt{b81}; $^{n}$ 
\citealt{b79}.} 
\end{table*}

Conclusive identification of the primary as a black hole (M$_1 >$ 3
M$_{\sun}$, \citealt{b110}) requires further observations once the
system has returned to quiescence. The following 3 quantities are
required to measure the mass of the primary: (i) radial velocity of the
secondary star, K$_2$ (ii) mass ratio, q $\rm \equiv \frac{M_2}{M_1}$, and
(iii) orbital inclination, i \citep{b114}. The first two measurements
can be made via  optical spectroscopy in quiescence. Measurement of the
third quantity is more complicated: in most cases one must model the
ellipsoidal modulation of the gravitationally distorted secondary
star. Crucially, this modelling requires that one must account for any
non-stellar flux present in the lightcurve.  

Observations of XRNe have predominately been carried out in the
optical. In this wavelength range (3000\AA $~\lesssim \lambda
\lesssim$ 9000\AA), the accretion disc is known to contribute
significantly to the observed flux. In quiescence, prominent emission
lines from the disc are observed (i.e. $\rm H_{\alpha}, H_{\beta}, HeI$
etc.) superposed on the spectrum of the secondary star. Doppler imaging
of XRNe in quiescence shows that the accretion disc is still present in
this state (e.g. \citealt{b86}), and reveals the presence of the
characteristic hotpot where the accretion stream impacts the
disc. Optical photometry also reveals the lightcurve of quiescent XRNe
to contain significant variability e.g. \citet{b117,b116,b113}.
Hence, these observations show that mass transfer from the
secondary star is continuously taking place. The contribution from the
hot accretion disc is typically observed to decrease from the U-band
($\sim$ 3650 \AA) to the R-band ($\sim$ 6500 \AA). As such it has been
assumed that the accretion disc should contribute even less at near
infrared wavelengths (NIR : 1 -- 2.5$\mu m$). Initial NIR spectroscopy
of GS 2023+338 and A0620-00 appeared to confirm this (\citealt{b18,b19},
although see \citealt{b36}). This appeared to justify the use of NIR
observations, rather than optical, in determining the ellipsoidal
variability (and constraining q \& i).     

However, recently it has become apparent that this might not always be
the case. \citet{b21} analysed quiescent K-band photometry of
GRO J0422+32. They found the NIR light curve to be dominated by a
flickering component, attributed to emission from the accretion disc, and
not an ellipsoidal modulation as one would expect if the secondary star
was the dominant source of the observed NIR flux. Using this data, the
mass of the black hole in this binary was estimated to be $\gtrsim$
10.4 M$_{\sun}$. 

Previously, \citet{b3} had measured the mass of the black hole
in this system to be 3.97 $\pm$ 0.95 M$_{\sun}$; crucially they assumed the
contribution of the accretion disc to be \textit{negligible}. 
A similar result was found in an investigation of the NIR flux in the
proto-typical XRN A0620-00 by \citet{b20}. Using moderate S/N spectra, they
measure the contribution of the secondary star to the observed flux to
be $82 \pm 2 \%$ in the H-band ($\sim$ 1.65$\mu m$). From this they
constrain the mass of the black hole to be M$\rm_x$ = 9.7 $\pm$ 0.6
M$_{\sun}$. In contrast \citet{b9} measured the black hole to have a mass
of 11 $\pm$ 1.9 M$_{\sun}$, and in this case the secondary was again
assumed to contribute essentially 100\% of the NIR flux. A number of
other black hole mass estimates have been made in the NIR (see Table
\ref{extinct-table}) and it is possible that the mass estimates for
these systems may also be subject to additional uncertainty. 

The presence of such a contribution from a cool accretion
disc component is expected theoretically; however, observational
evidence is difficult obtain \citep{b7}. In addition, at longer
wavelengths (i.e. MIR: 3 -- 24$\mu m$), \textit{Spitzer} observations
of a number of XRNe indicate the presence of a circumbinary disc which
will contribute significantly at these wavelengths \citep{b6}. 

These results have lead us to re-examine the issue of the non-stellar
contribution to the NIR flux of quiescent XRNe and in particular to
re-evaluate the degree to which NIR studies improve our chances of
constraining \textit{q, i,} and hence the black hole mass, in comparison
to optical studies. These results are presented herein as follows: we
introduce our sample systems and the relevant data in \S2, and show the
relevant spectral energy distributions (SEDs) in \S3. It is immediately
apparent that there is a significant non-stellar contribution to the
observed flux in the NIR. Modelling the multi-wavelength SEDs shows that
this excess is consistent with blackbody emission from a combination of
the accretion disc and a circumbinary disc. A discussion of our results
follows in \S4.

\section{Data}

Our sample consists of XRNe with quiescent data spanning the optical to
at least the NIR wavelength range. In addition, we choose only those
systems that contain main sequence companions (class V), with the exception
of GS 2023+338. This XRN has a much longer orbital period and a sub-giant
companion star (class IV). However, as this is the only other quiescent
XRN with coverage at \textit{Spitzer} wavelengths, it has been included
for completeness. The XRNe that make up our sample include 6 suspected
to contain a black hole primary and a single neutron star system. In
Table \ref{extinct-table}, we list the primary system parameters for each
binary. Our data consists of archival published optical and NIR
photometry as well as a number of mid-IR (MIR: 3 -- 24$\mu m$) data
points from the \textit{Spitzer} observatory \citep{b6}. In particular
\textit{Spitzer} photometry exists for 4 of the systems, namely:
A0620-00, XTE J1118+480, GS 2023+338 \& Cen X-4. We have also recently
obtained a medium resolution K-band spectrum of the black hole system
GS 2000+25, which we discuss below. In Table \ref{sed-table}, we list the
dereddened flux from each system in the wavelength regions of interest.

\subsection{GS 2000+25 Spectroscopy}
K-band spectra of GS 2000+25 were obtained with the Near-IR imager (NIRI,
\citealt{b75}) using the f/6 camera at the Gemini-North telescope. 
Medium-resolution spectra of GS 2000+25 were obtained on the nights of
2006 August 23, 25 and September 1 UT as part of the Gemini service
program (Prog. id: GN-2006B-Q-85, PI: Robinson). The K grism
(1.9 -- 2.49$\mu m$) was used in conjunction with a 0.23 $\arcsec$ slit,
giving a resolution R $\approx$ 1100. Individual exposure times were
300s, dithered to 5 positions along the slit. In total 37 spectra of the
system were acquired yielding a total exposure time of $\sim$ 3hrs.
At the time of our observations photometry revealed the K-band magnitude
of GS 2000+25 to be 16.6 $\pm$ 0.15, where the error is dominated by the
error in the 2MASS\footnote{The Two Micron All Sky Survey is a joint project
of the University of Massachusetts and the Infrared Processing
and Analysis Center, California Institute of Technology, funded
by NASA and the National Science Foundation.} stars relative to which
our image is calibrated. This is consistent with previous measurements
of the system, showing it to be in quiescence at the time of our
observations.  

The spectra were observed in 3 separate groups of 23, 8 and 6 science
frames taken on each of the three nights outlined above. The airmass was
typically in the range 0.95 -- 1.15 during the observations as such
slitlosses due to atmospheric differential refraction were negligible.
The one-dimensional spectra were extracted within
  \textsc{iraf}\footnote{\textsc{iraf} is distributed by the National Optical
  Astronomy Observatories, which are operated by the Association of
  Universities for Research in Astronomy Inc., under cooperative
  agreement with the National Science Foundation.} using both the
\textsc{noao.onedspec} and the \textsc{gemini.niri} packages. Both
sets of extracted spectra were found to be consistent. Unfortunately,
due to the nature of the service observations a significant fraction
of the spectra were taken in poor to deteriorating conditions
resulting in a large number of unusable exposures. Hence the final
signal to noise ratio (S/N) of our science spectrum was lower than
anticipated. Spectra of the A0V stars HD182761 and HD197291 were also
taken before and after the science frames respectively to aid with the
removal of the telluric features and flux calibration of our science
spectrum.   

\begin{figure}
\includegraphics[height=84mm,width=64mm,angle=-90]{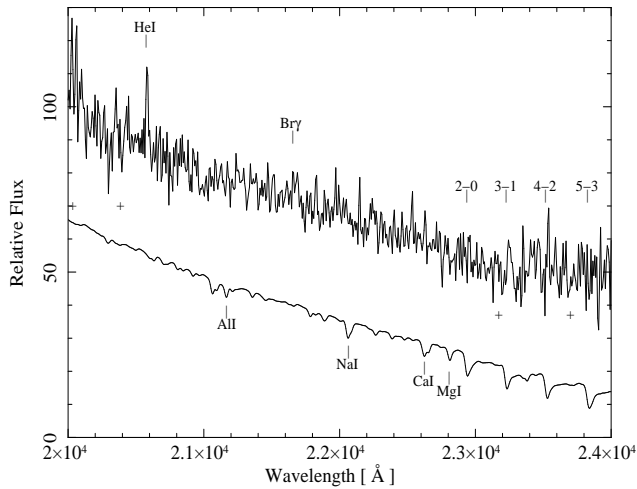} 
\caption{Gemini NIRI K-band spectrum of GS 2000+25 (top), the HeI
  and Br$\gamma$ emission lines are indicated, also indicated are the
  positions of prominent atomic absorption lines and the
  CO-bandheads. The crosses mark the position of telluric residuals, the
  spectrum of a K5V star is plotted underneath for comparison.} 
\label{gs2000spec}
\end{figure}

Correction of the science frames for telluric absorption was carried out
following the prescription of \citet{b76}. The low S/N of the individual
spectra complicated the removal of the telluric features, with a number
of residuals remaining. The resulting spectra were corrected for the
Doppler shift of the spectral lines induced by the orbital motion of
the secondary star using the ephemeris of \citet{b47} and median
combined so as to negate the effect of cosmic-ray hits and other
spurious effects, while at the same time maximising the available
S/N. This spectrum was then dereddened using the extinction listed in
Table \ref{extinct-table}. As the slit losses are large, only a relative
flux calibration was possible. The resulting spectrum ($\sim$ 2hrs
exposure) is displayed in Fig. \ref{gs2000spec}; emission lines of HeI
and Br${\gamma}$ are observed.

\begin{table*}
\caption{The spectral energy distribution data. All fluxes are units of
  erg s$^{-1}$ cm$^{-2}$.} 
\label{sed-table}
\begin{center}
\begin{tabular}{lcccccccccccc}
\hline
Source & B & V & R & I & J & H & K & 4.5$\mu m$ & 8$\mu m$ & 24$\mu m$ &
Radio\\ 
& 4400\AA& 5500\AA & 6400\AA & 7900\AA & 1.25$\mu m$ & 1.65$\mu m$ &
2.2$\mu m$ & & & & 8.5GHz\\ 
\hline\hline 
XTE J1118+480 $^{1, 2}$ & 2.54e-13 & 2.89e-13 & 2.94e-13 & -- &
3.50e-13 & 2.78e-13 & 2.1e-13 & 3.07e-14 & 1.69e-14 & $>$ 2e-15 & -- \\
GRO J0422+32 $^{3}$ & -- & 7.35e-14 & 1.18e-13 & 1.39e-13 & 2.32e-13 & 2.07e-13
& 1.07e-13 & -- & -- & -- & -- \\
A0620-00 $^{4, 5, 6, 7, 2}$ & 1.23e-12 & 2.09e-12 & 2.30e-12 & 2.38e-12 &
3.25e-12 & 2.88e-12 & 1.58e-12 & 2.99e-13 & 9.34e-14 & 6.75e-15 &
4.33e-18\\ 
GS 2000+25 $^{8, 9}$ & -- & -- & 8.34e-13 & 6.34e-13 & 7.70e-13 & 7.52e-13 &
3.53e-13 & -- & -- & -- & --\\
GS 1124-683 $^{10}$ & 8.98e-14 & 2.20e-13 & 2.22e-13 & 2.49e-13 &
2.90e-13 & 2.98e-13 & 1.43e-13 & -- & -- & -- & --\\ 
Cen X-4 $^{11, 2}$ & 6.15e-13 & 9.09e-13 & 1.39e-12 & 1.59e-12 & 2.08e-12 &
1.84e-12 & 1.11e-12 & 1.33e-13 & 3.56e-14 & $>$ 3.8e-15 & --\\  
GS 2023+338 $^{12, 13, 2}$ & 6.88e-12 & 1.54e-11 & 2.34e-11 & -- & 3.11e-11 &
2.57e-11 & 1.29e-11 & 2.01e-12 & 5.44e-13 & 1.91e-14 & $\sim$ 4e-17 \\ 
\hline
\end{tabular}
\end{center}
\medskip
{
REFERENCES: $^{1}$ \citealt{b5};  $^{2}$ \citealt{b6}; $^{3}$
\citealt{b3}; $^{4}$ \citealt{b9}; $^{5}$ \citealt{b36}; $^{6}$
\citealt{b98}; $^{7}$ \citealt{b10}; $^{8}$ \citealt{b11}; $^{9}$
\citealt{b12} ; $^{10}$ \citealt{b13}; $^{11}$ \citealt{b66}; $^{12}$
\citealt{b79}; $^{13}$ \citealt{b99}.}
\end{table*}


\section{Analysis}

\subsection{GS 2000+25 K-band Spectrum}\label{gs2000spectrum}
The K-band spectrum in Fig. \ref{gs2000spec} is displayed at a
resolution of $\sim$ 1100 (270 km s$^{-1}$) and has a S/N ratio of
$\sim$ 10 at 2.3$\mu m$. We observe emission features consistent with
both HeI and Br$\gamma$ with equivalent widths of $\sim$ 6\AA~\& 
12\AA~respectively. Also indicated are the positions of a number of the
atomic absorption features expected from the secondary star.

We also attempted to place on upper limit on the presence of any CO
bandheads in the GS 2000+25 spectrum. As the S/N degrades further at longer
wavelengths, our efforts were confined to the CO (2--0) band head at
2.294$\mu m$. We used a model of this bandhead in the K5V star spectrum
to place an upper limit on its presence in the observed GS 2000+25
spectrum. We can rule out the presence of this bandhead with an
equivalent width of $\geq$ 20\% of that expected from a K5V star (solar
abundance) at the 99\% confidence level. 

The spectral slope is observed to be consistent with that expected
from the K5V secondary star in this system for $\lambda \leq$ 2.3$\mu
m$: for longer wavelengths, the slope appears to flatten. In an effort
to constrain the amount of non-stellar flux present a number of spectra
were simulated. Noise was added to a K5V spectrum consistent with the
secondary in this system until a S/N ratio similar to the science
spectrum was achieved. To this a constant contribution representing an
accretion disc with a temperature profile of the form T $\propto$
r$^{-1/2}$ was added (see \S \ref{results}). We find for a fractional disc
contribution of 25\%, the simulated spectral slope is no longer
consistent with the observed spectrum, indicating a likely disc
contribution of $\lesssim$ 25\% in the 2.0 -- 2.3$\mu m$
wavelength region and more at wavelengths greater than this.

\subsection{Multiwavelength Spectral Energy Distributions}\label{m-seds}
In Figures \ref{the-seds1} \& \ref{the-seds2}, we plot the extinction
corrected optical/IR SEDs of the XRNe as listed in Table
\ref{sed-table}, where the error bars account for the uncertainty in the
photometry and the reddening. Here, we fit model atmospheres of the
appropriate spectral type to the SEDs of the quiescent XRNe: these data
include optical, NIR and \textit{Spitzer} measurements. In contrast to
previous authors who normalised to the H-band \citep{b5} or K-band flux
\citep{b6,b3,b13}, where the IR contamination from the accretion disc
was assumed to be minimal, we choose instead to normalise these models
relative to the measured secondary contribution in the R-band, which has
generally been more accurately determined using optical
spectroscopy. The models we fit are NextGen model atmospheres
\citep{b17,b16} corresponding to the spectral type of the secondary star
in each system. We restrict ourselves to those veiling measurements for
which the simultaneously measured R-band flux is consistent with the
value used in the SED. The veiling in all cases is defined as a
percentage of the total system flux emitted by the accretion disc,
typically measured in the region of the H$\alpha$ line (6562\AA):
$$\rm F_{veil} = \frac{F_{disc}}{F_{star} + F_{disc}} $$ and is indicated in
the top left hand corner of each SED in Fig. \ref{the-seds1} \&
\ref{the-seds2} and in the final column of Table \ref{extinct-table}. As
the various photometric data points used were non-simultaneous, we do
not attempt to rigorously quantify the excess in the NIR. The quiescent
magnitudes of XRNe are typically observed to vary by $\sim$ 0.1
magnitudes (e.g. \citealt{b116}), in addition to the $\sim$ 0.2 magnitude
periodic variability due to the intrinsic ellipsoidal modulation of the
secondary star. However, when normalised this way, the SEDs of these
XRNe are all seen to exhibit an excess of flux at NIR wavelengths
comparable to that measured in the optical (also indicated in Figures
\ref{the-seds1} \& \ref{the-seds2}). In this sense the NIR appears to be
just as contaminated as the optical, in the context of measuring the
flux from the secondary. 

\subsection{A Simple Model}\label{a-simple-model}
The current paradigm for understanding the emission from quiescent
XRNe involves a thin disc which transitions to a quasi
spherical inner flow at a distance of $\sim$ 10$^3$ -- 10$^4$
Schwarzschild radii from the compact object. The inner flow is thought
to consist of an advection dominated accretion flow (see
\citealt{b120} and references therein), although it has also been argued
that the inner region instead consists of a jet/outflow (\citealt{b89,c96}). 
In an effort to place a constraint on the source of the observed excess
NIR flux, we have attempted to model the observed SED. The spectral
energy distributions are fit with a multi-component model consisting of:   
\begin{enumerate}
\item{A spherical blackbody representing the emission from the
  secondary star.}
\item{An accretion disc.\\}
The flux from the accretion disc may be modelled in the form of a multi-colour
blackbody where the temperature profile across the surface of the
optically thick geometrically thin accretion disc is \citep{b15}:
\begin{equation}
\rm T_{eff}(r) = \left(\frac{3GM_1\dot{m}}{8 \pi r^3 \sigma}
\left[1-\sqrt{\frac{R_{in}}{r}}\right]\right)^{1/4}  
\end{equation}
This results in T$\rm_{eff}(r) \propto r^{-3/4} for ~r \gg R_{in}$ in the
case of a viscously heated steady state disc. 

Alternatively, we might
expect a flatter temperature profile: for example, T$\rm_{eff}(r) \propto
r^{-1/2}$ has been observed in the quiescent accretion disc in some
cataclysmic variables (CVs, \citealt{b108,b107,b103}). Such a temperature
profile is intermediate between the classic steady state viscously
heated accretion disc (r$^{-0.75}$) and the irradiated disc case
(r$^{-0.43}$).
\item{A circumbinary disc.\\}
If we assume the circumbinary disc to be opaque, flat and passively
illuminated by the central star \citep{b14}, we expect  the following
temperature profile:
\begin{equation}
\rm T_{eff}(r) = \left(\frac{2}{3\pi}\right)^{1/4}T_*
\left[\frac{R_{*}}{r}\right]^{3/4}  
\end{equation}
where R$_*$ \& T$_*$ are the radius and temperature of the irradiating star
respectively.\\ 
\end{enumerate}

\subsection{Results}\label{results}

Given the non-simultaneous nature of the observations we do not try to
rigorously fit the data (e.g. via $\chi^2$ minimisation). None the less,
in each case it is clear that a model consisting of an accretion disc
extending from 10$^3$ -- 10$^5$ Schwarzschild radii (R$\rm_S$) to an outer
radius of $\sim$ 0.5 the Roche lobe radius of the compact object
(R$\rm_{L1}$), in addition to the secondary star, is required to provide
the observed NIR flux. This 
is consistent with previous observations of the accretion disc in
quiescence \citep{b88,b87,b109,b86}. At \textit{Spitzer} wavelengths,
the flux appears to be a combination of emission from the accretion
disc/circumbinary disc. The SEDs of all the XRNe are seen
to follow this general pattern. We emphasise that although the
following analysis is qualitatively similar to that carried out by
\citet{b6}, it differs in the important aspect that we normalise to the
R-band flux, which results in a significant non-stellar component to
the observed NIR flux (see \S \ref{m-seds}). We discuss the fits to
each system in more detail below.     

\subsubsection{A0620-00}
\begin{figure*}
\subfigure[\textbf{A0620-00}]{\includegraphics[height=0.48\textwidth,angle=-90]{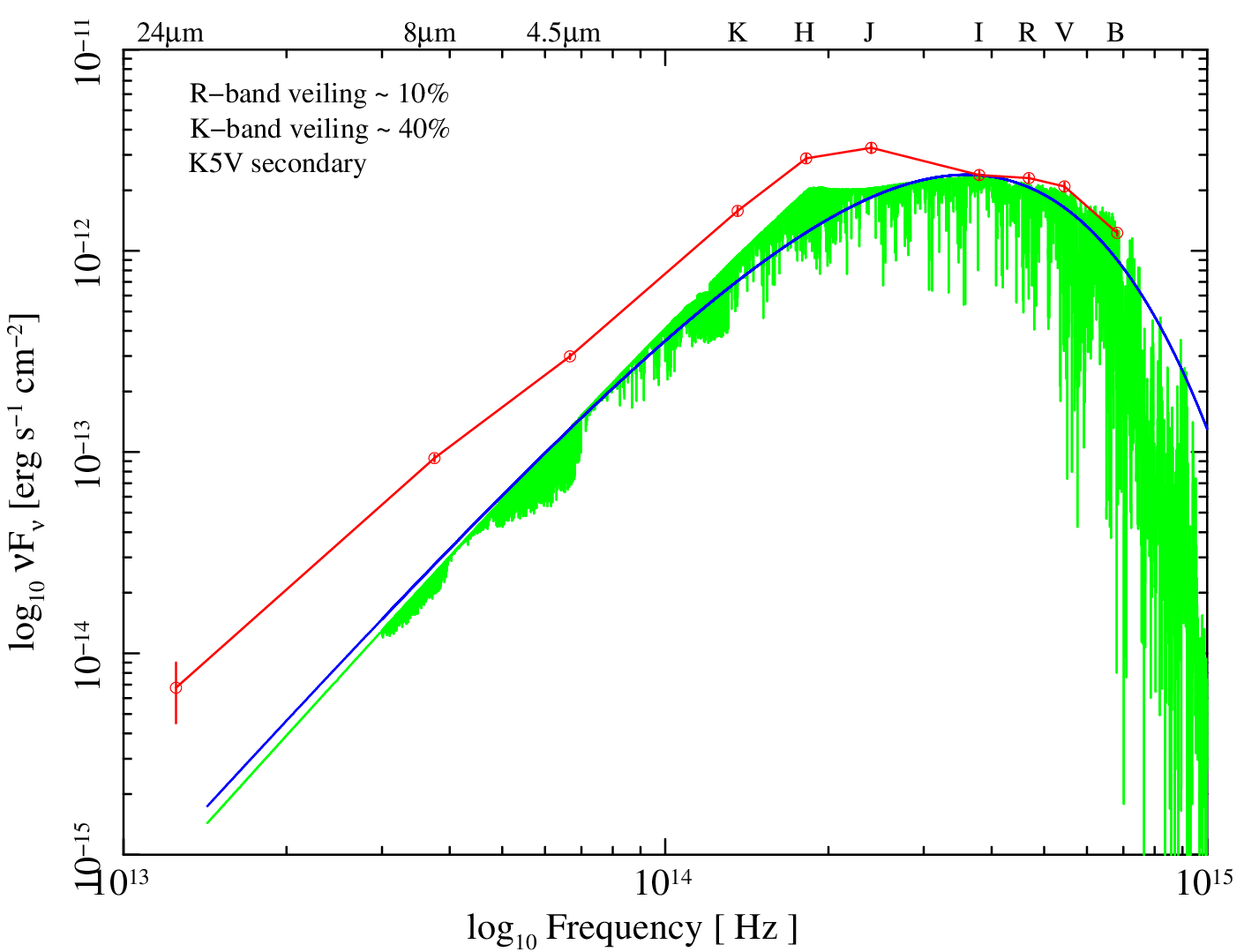}}   
\subfigure[\textbf{XTE J1118+480}]{\includegraphics[height=0.48\textwidth,angle=-90]{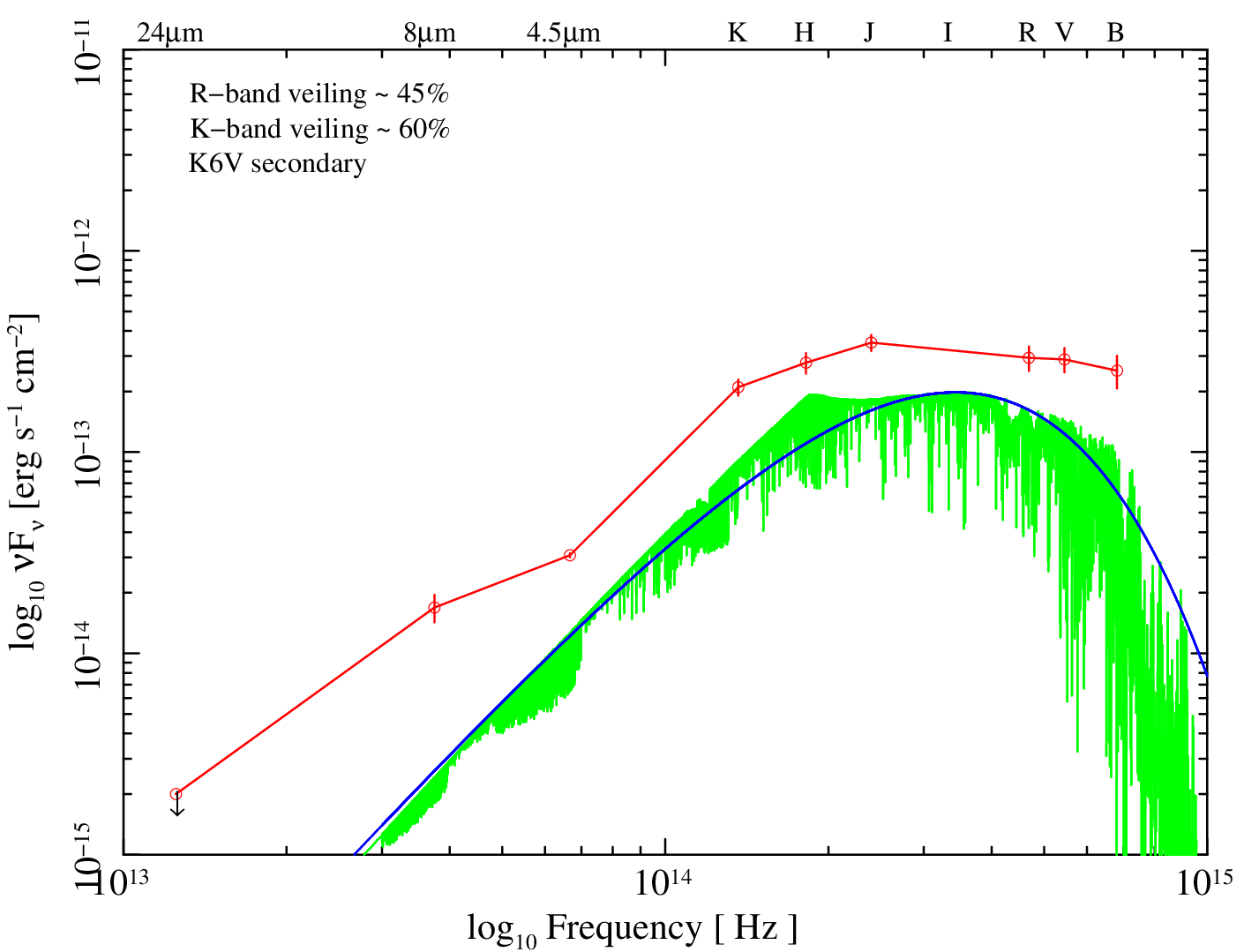}} 
\subfigure[\textbf{GS 2000+25}]{\includegraphics[height=0.48\textwidth,angle=-90]{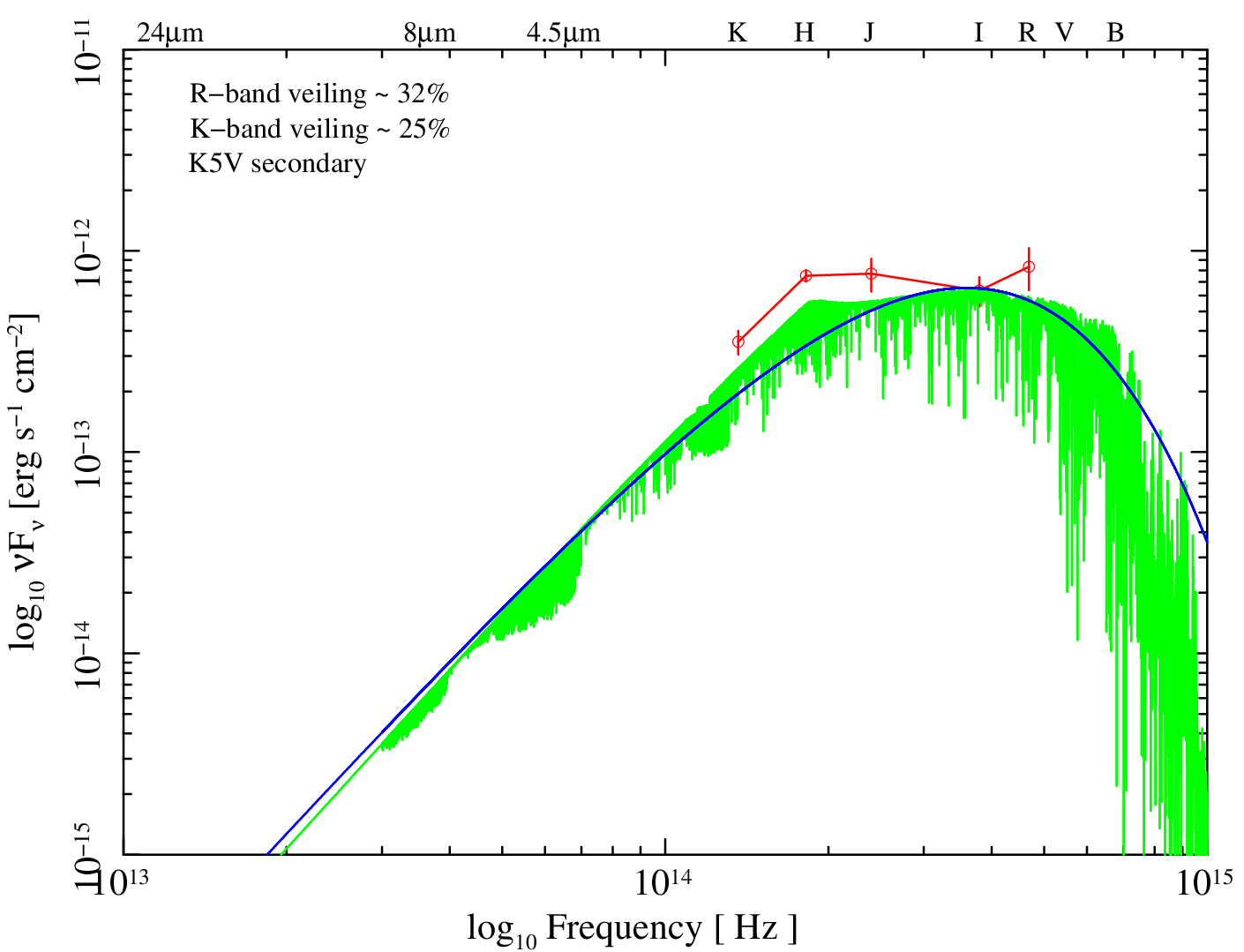}} 
\subfigure[\textbf{GRO J0422+32}]{\includegraphics[height=0.48\textwidth,angle=-90]{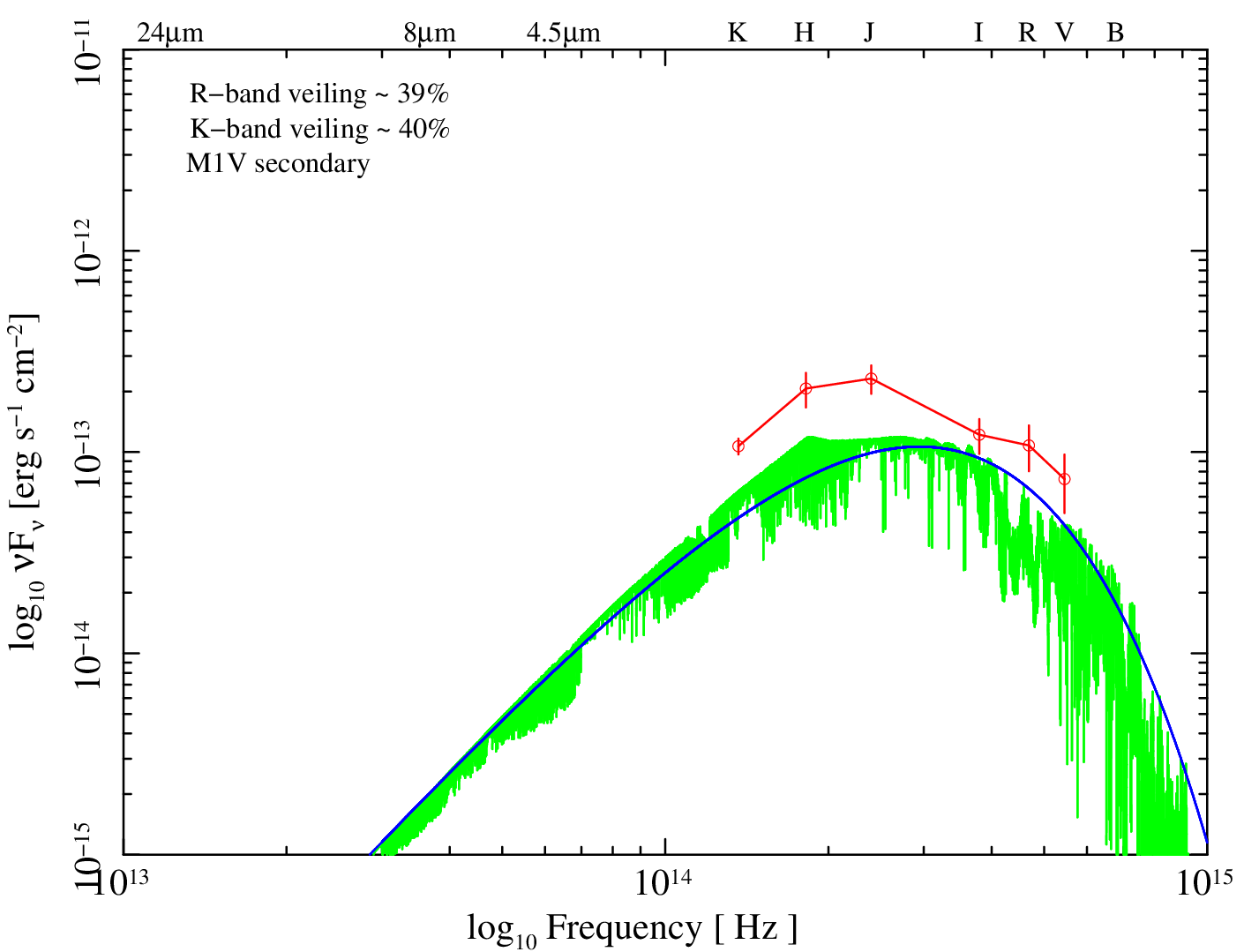}}
\caption{Optical/IR SEDs for the XRNe outlined in the text. The grey
  curve is the absorption corrected SED, where the error bars account for
  the uncertainty in the photometry and reddening. The black curve
  represents the blackbody corresponding to the temperature of the model
  atmosphere (light-grey). The R-band accretion disc contamination is listed
  (see Table \ref{extinct-table}) along with the resulting K-band
  contamination and the adopted spectral type of the secondary star in
  the top left.}  
\label{the-seds1}
\end{figure*}

\begin{figure*}
\subfigure[Steady state disc
  model, T $\propto$ r$^{-0.75}$]{\includegraphics[width=0.37\textwidth,angle=-90]{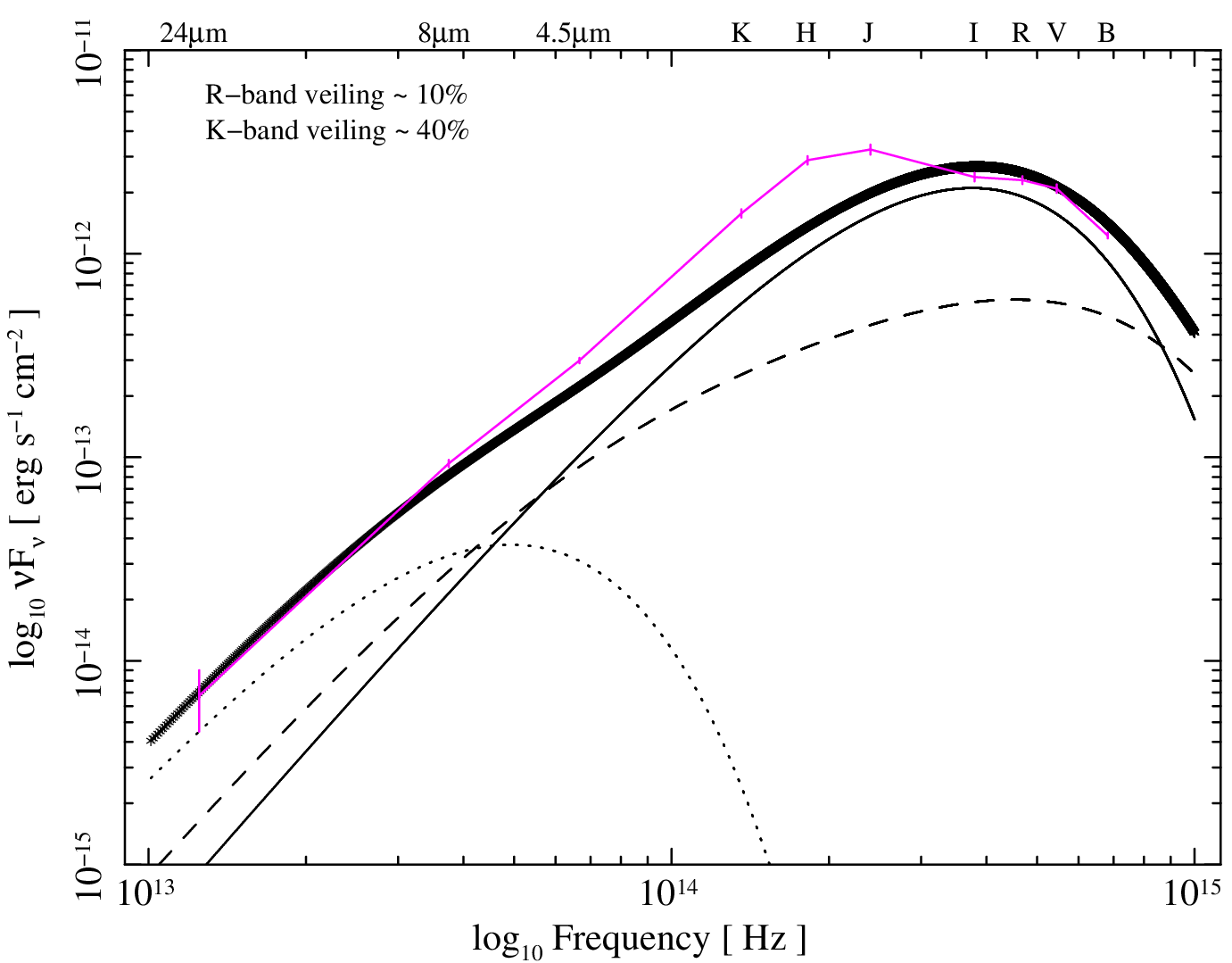}} 
\subfigure[Flat temperature profile disc
  model, T $\propto$ r$^{-0.5}$]{\includegraphics[width=0.37\textwidth,angle=-90]{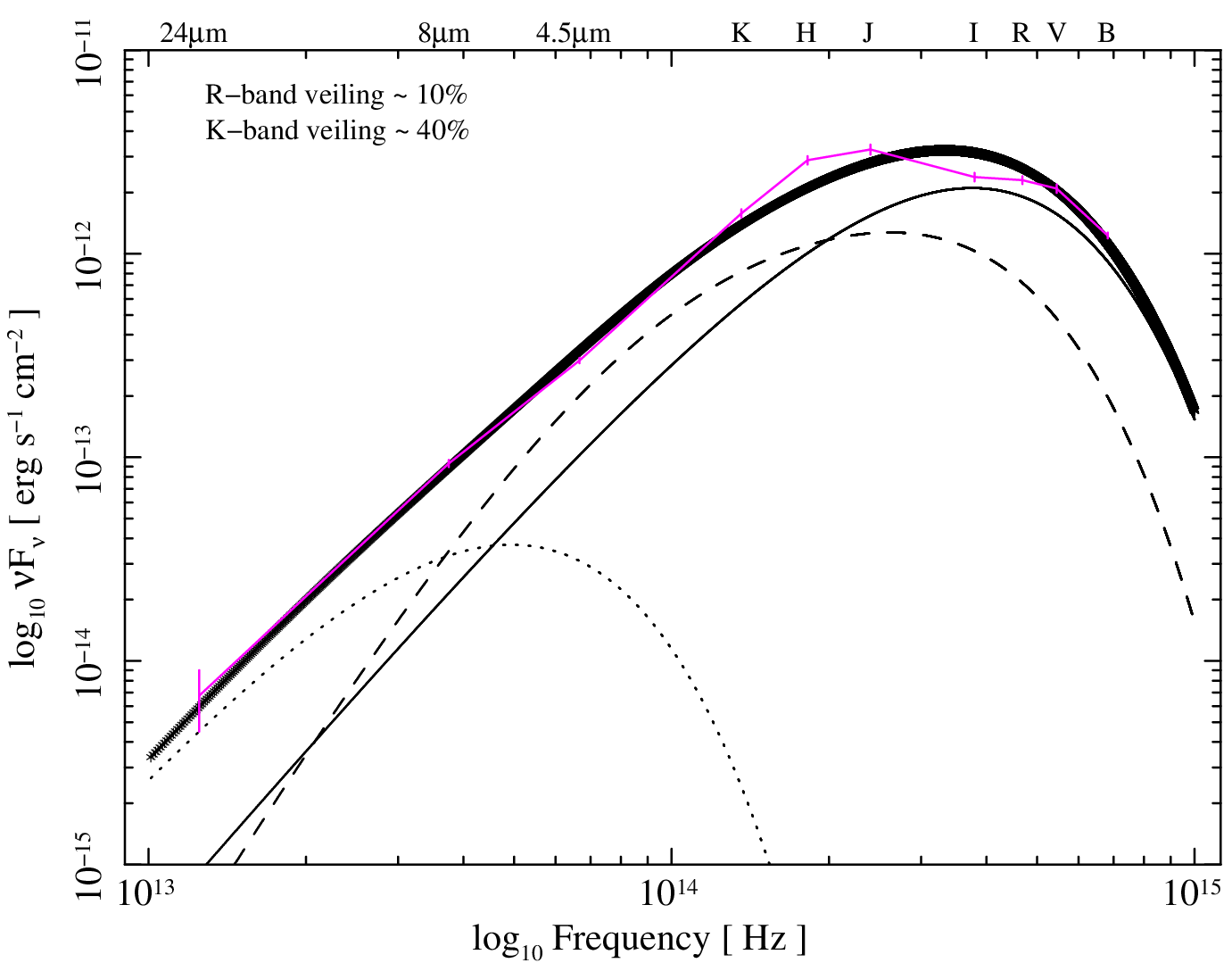}} 
\caption{Blackbody fit to the A0620-00 SED consisting of (i) star -
  solid line (ii) multi-colour blackbody disc - dashed line (iii)
  circumbinary-disc - dotted line. The observed SED is in grey with
  the model SED indicated by the thick line.}
\label{a0620-bbfit}
\end{figure*}

In Figure \ref{a0620-bbfit}, we display the resulting fit in the case
of A0620-00. A steady state disc fit is displayed on the left; a
disc with a flatter temperature profile (T $\rm \propto r^{-0.5}$) is
displayed on the right. We see that the disc with the flatter
temperature profile provides a better fit to the observed SED. The basic
model consists of a blackbody at the temperature of the model
atmosphere displayed in Figure \ref{the-seds1}, a 'flat' accretion disc
extending from $\sim$ 10$^4$ R$\rm_S$ to $\sim$ 0.5 R$\rm_{L1}$ and a
circumbinary disc consistent with the previous estimates of \citet{b6}. 
Such an accretion disc will contribute $\sim$ 40\% of the K-band flux.     
In comparison, \citet{b20} recently determined the accretion disc to be
contributing $\sim$ 23\% of the K-band flux through moderate S/N
spectroscopy. Given the uncertainties discussed elsewhere (\S 3.2, \S
4), we regard these values to be consistent with each other. 

\begin{figure*}
\subfigure[\textbf{GS 1124-683}]{\includegraphics[height=0.48\textwidth,angle=-90]{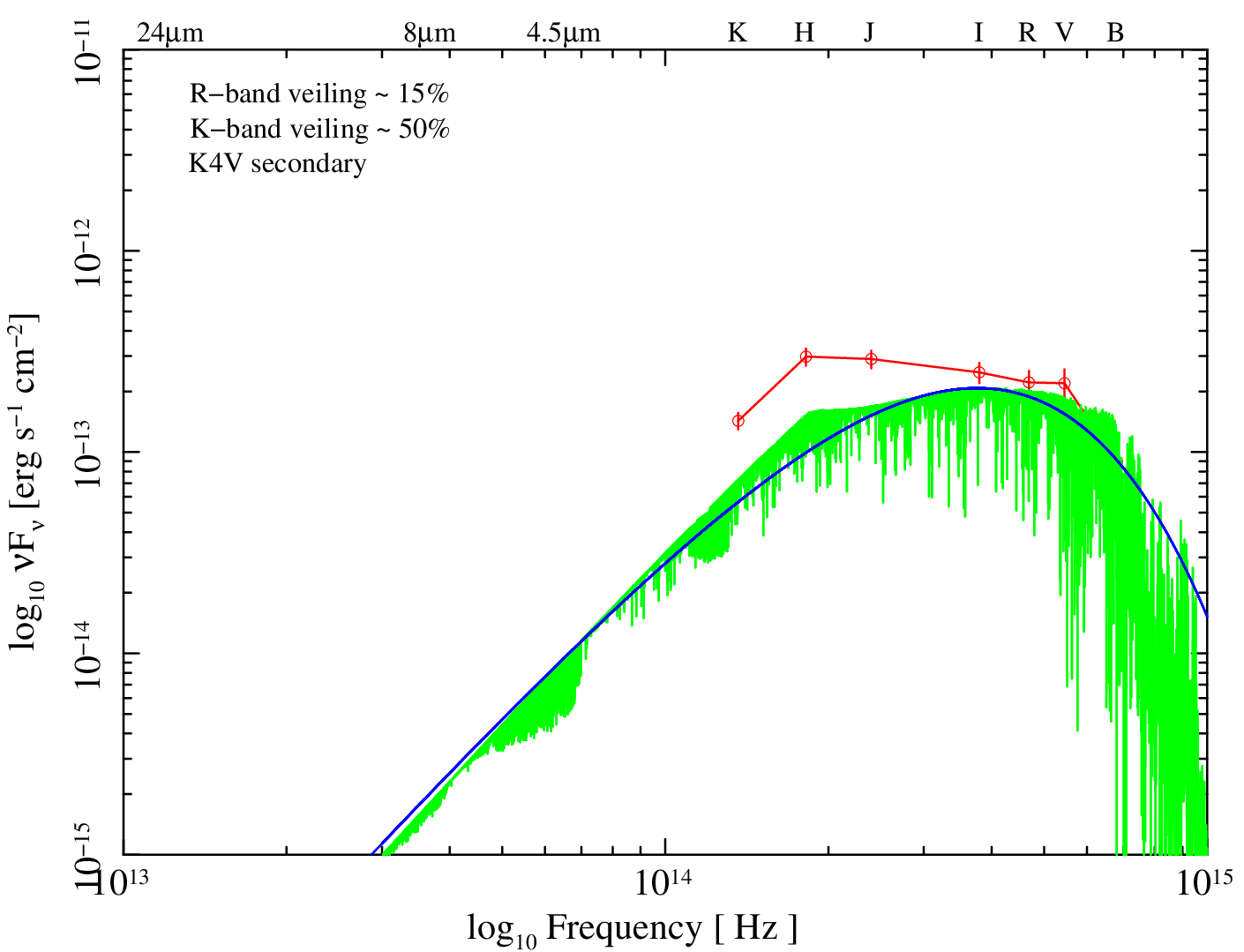}} 
\subfigure[\textbf{GS 2023+338}]{\includegraphics[height=0.48\textwidth,angle=-90]{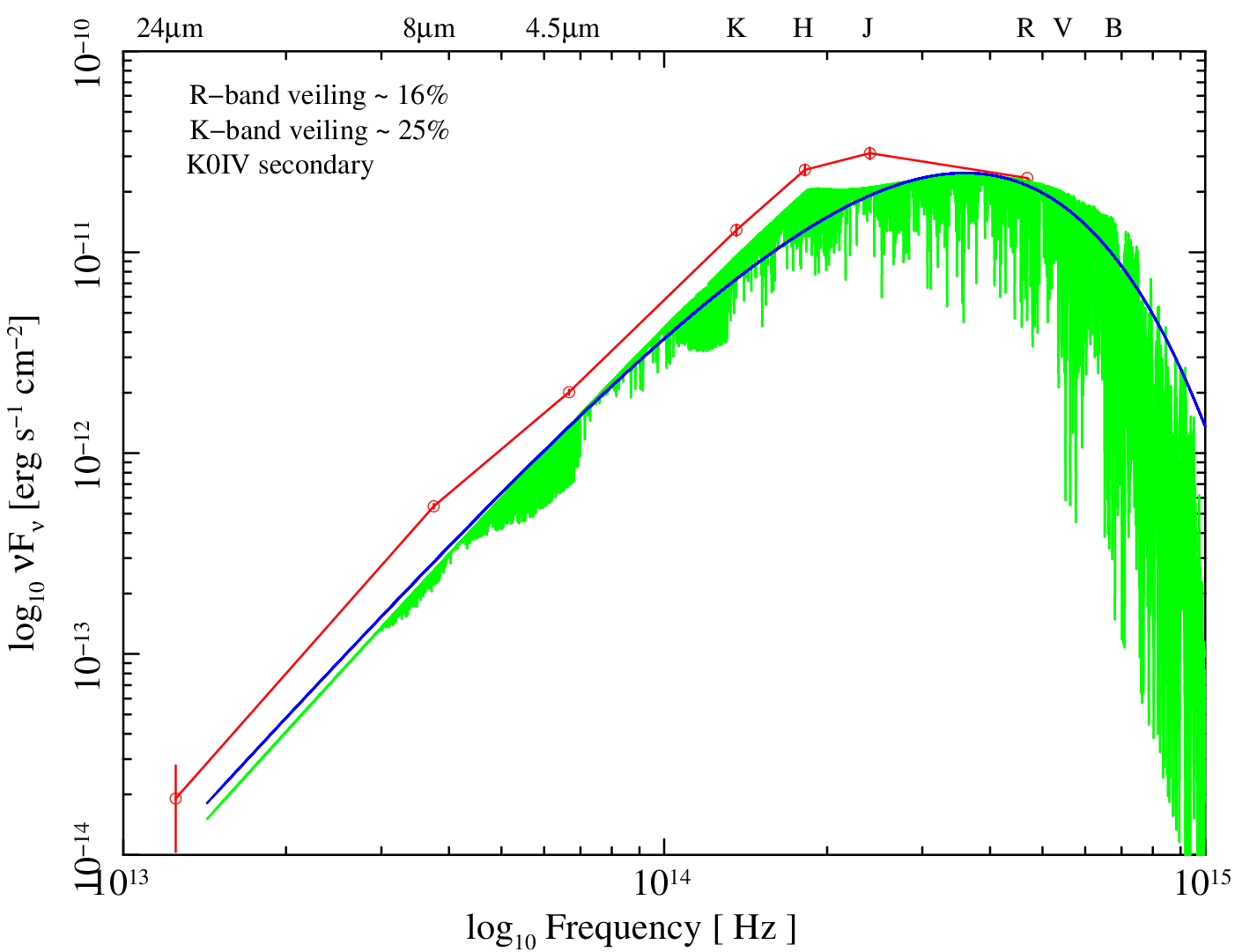}}  
\subfigure[\textbf{Cen X-4}]{\includegraphics[height=0.48\textwidth,angle=-90]{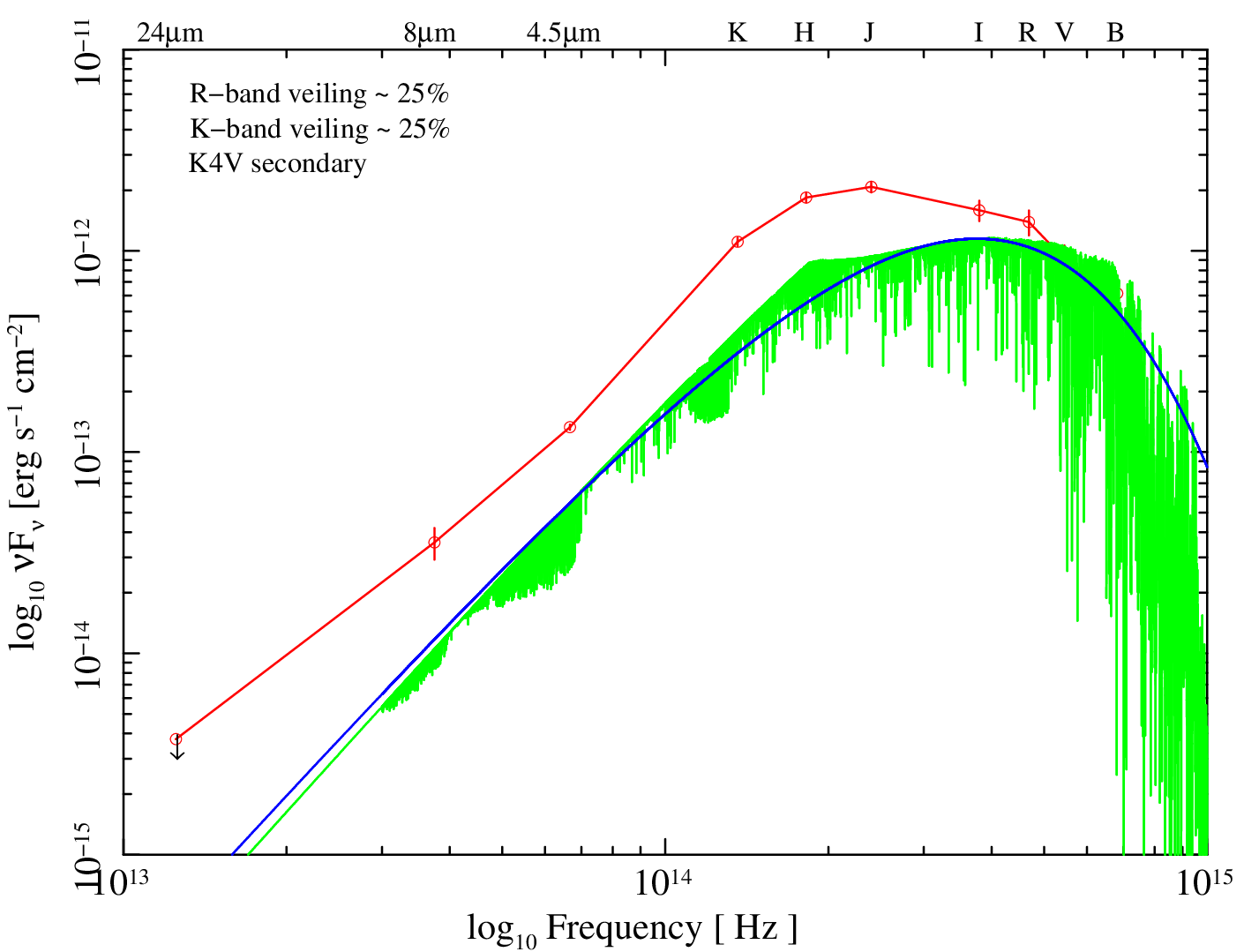}}  
\caption{As in Figure \ref{the-seds1}.}  
\label{the-seds2}
\end{figure*}

\subsubsection{XTE J1118+480}
Here we find a similar model to that described above, in particular a
flat temperature profile accretion disc is required. The slope of the
SED is consistent with emission from the secondary plus the accretion
disc out to 4.5$\mu m$, however the 4.5 - 8$\mu m$ slope clearly
indicates the presence of an additional source of flux. 

In contrast to \citet{b6}, we find a standard circumbinary disc is
unable to account for the observed excess. This is because we account for
the accretion disc veiling, which results in a lower contribution from
the secondary star at \textit{Spitzer} wavelengths in this work, 
requiring a greater flux contribution from the circumbinary disc.
Instead, we find a circumbinary disc with a flatter temperature
profile similar to the accretion disc is required. In this case the
accretion disc will contribute $\sim$ 60$\%$ of the flux at NIR
wavelengths.

\subsubsection{GS 2023+338}
As is immediately obvious from the SED in Fig. \ref{the-seds2}, the flux
from this system is dominated by the emission from the type IV secondary
star. We find the accretion disc to contribute in the region of 25\% of
the flux in the JHK region of the spectrum. 
The excess at 24 microns is
also consistent with an origin in the accretion disc. However, we are
unable to rule out a contribution from a circumbinary disc as suggested
by \citet{b6}.

\subsubsection{Cen X-4}
Cen X-4 is the only system in our sample where the compact object is a
neutron star. However, as in the other systems the secondary star is
unable to account for the NIR/optical flux, with there being a large
excess over that from the secondary at NIR and \textit{Spitzer}
wavelengths. The magnitude of the contribution from the accretion disc
is consistent with that observed in the black hole systems mentioned
above. The disc is required to extend from an inner radius of 0.3 R$_{\sun}$
to an outer radius of 0.5 R$\rm_{L1}$. We also note that the 8$\mu m$ flux
in conjunction with the 24$\mu m$ upper limit allows for the possible
presence of a circumbinary disc, which contributes up to 20$\%$ of the
8$\mu m$ flux, in contrast to the analysis of \citet{b6}. Such a
circumbinary disc will not contribute to the observed K-band emission.

\subsubsection{Systems Without \textit{Spitzer} Measurements}
\textit{GRO J0422+32:} Previous observations of GRO J0422+32 highlighted
the likely presence of a large non-stellar NIR component in this system
\citep{b21}. Modelling of the SED supports this claim, with the
accretion disc component providing $\sim$ 40\% of the observed NIR
flux.\\
\textit{GS 2000+25:} Again, here we find the accretion disc contribution
to be large, $\sim$ 30\% in the K-band.\\ 
\textit{GS 1124-683:} The flux from the accretion disc is seen to be
$\sim$ 40-50\% of the total flux at NIR wavelengths.  


\section{Discussion}
As has been demonstrated in the previous section, it is clear that the
NIR flux observed from quiescent X-ray novae contains a significant
component that does not originate in the secondary star. From an
analysis of the spectral energy distributions of these XRNe, it now
appears that this NIR excess is dominated by thermal emission from the
cooler outer regions of the accretion disc. This excess requires the
presence of a cool accretion disc with a temperature profile T $\propto$
r$^{-0.5}$; such a cool disc component (T $\sim$ 3000 - 4000 K) is
expected on theoretical grounds (e.g. \citealt{b7}). This temperature
profile is intermediate between the classic steady state viscously
heated accretion disc (r$^{-0.75}$) and the irradiated disc case
(r$^{-0.43}$). The use of a temperature profile of this form is
supported by the observation of similarly shallow temperature profiles
from eclipse mapping of the accretion discs in quiescent dwarf novae
(see \citealt{b103} and references therein). 

Given the difficulty in accurately determining the spectral type of the
secondary star in XRNe, we also investigated the effect of varying the
spectral type (within the range listed in Table \ref{extinct-table}) on
the magnitude of the NIR excess. In particular, a later spectral type
could account for a significant portion of the observed NIR excess
(Fig. \ref{the-seds1} \& \ref{the-seds2}). Using the latest secondary
star spectral types
allowed, as previously determined (see Table \ref{extinct-table}), the
analysis in \S \ref{results} was repeated. The observed NIR flux is
found to remain systematically greater than the relevant model
atmosphere for all the systems in our sample, although the magnitude of
the excess is found to decrease as expected. In the 7 systems in our
sample, the observed K-band excess will remain high ranging from $\sim$
10\% in A0620-00 (K7V secondary) to 40\% in XTE J1118+480 (M0V
secondary). The veiling in each system is typically known to within $\pm$
5\%, which  can lead to an error in the disc flux ranging from 50\%
(A0620-00: 10 $\pm$ 5) to as little as 12\% (Cen X-4: 25 $\pm$ 3). We adopt
this as an appropriate estimate for the accuracy of our K-band veiling
measurements. Regardless of the magnitude of the excess, it is the
\textit{systematic presence} of this excess in all of the surveyed systems that
is the most convincing evidence for its existence.  

In the following sections, we discuss additional spectroscopic evidence for
accretion disc contamination of the NIR flux in XRNe, which support the
SED analysis given in \S \ref{results} and above. In addition, we
consider possible jet emission in the NIR, briefly discuss the MIR flux
and compare the systems containing black hole primaries to the neutron
star system. We end by considering the implications of neglecting to
properly account for this additional flux when determining black hole masses.  

\subsection{Previous K-band Spectroscopy}
NIR spectroscopy of quiescent black hole XRNe has also provided evidence
for a non-negligible contribution from the accretion disc. Recent
moderate resolution observations of A0620-00 by \citet{b20}, support the
presence of a near-infrared excess in this system. Here the spectral
slope was observed to deviate significantly from that expected of the
secondary star alone. Detailed analysis showed the secondary star to
contribute 82 $\pm$ 2\% of the H-band flux and $\sim$ 77\% of the
K-band flux. \citet{b21} analysed low resolution spectra of the XRN GRO
J0422+32 that displayed significant Br$\gamma$ emission; in combination
with photometric observations they showed the accretion disc to be
contributing up to $\sim$ 30\% or more of the observed K-band flux.

\subsection{The GS 2000+25 Spectrum}
In this work, we have obtained a medium resolution K-band spectrum of GS
2000+25. Emission lines consistent with the accretion disc are detected
(HeI, Br$\gamma$), whereas the intrinsic absorption lines of the
secondary star appear to be absent (although the S/N is low, see
Fig. \ref{gs2000spec}). Modelling of the CO (2--0) bandhead at 2.294$\mu
m$ allows us to place a limit on the equivalent width of this feature in
our spectrum. We can exclude the presence of this bandhead with an
equivalent width of $\geq$ 20\% of that expected from a K5V star, at the
99\% confidence level. On the other hand, analysis of the spectral slope
of our spectrum allows us to conservatively constrain the accretion disc
to be contributing $\lesssim$ 25\% of the flux for $\lambda \leq$
2.3$\mu m$, and more at wavelengths greater than this, although a more
accurate determination is limited by our low S/N ratio. We regard this
as consistent with the value estimated in \S \ref{m-seds} \&
Fig. \ref{the-seds1}. 

These results indicate that the CO abundance in the secondary star in GS
2000+25 is anomalous. If the CO abundance were similar to solar, we
would expect to detect the CO (2--0) bandhead at approximately 75\% of
the equivalent width found in the K5V spectrum, with the remainder of
the line being filled in by the accretion disc continuum, which we have
shown contributes $\lesssim$ 25\% of the observed flux. Instead, we
limit the maximum CO (2--0) bandhead strength to be 20\% of that
expected from a K5V star at solar abundance. 

In recent observations of A0620-00 at similar resolution
(albeit higher S/N) both \citet{b20} and \citet{b104} detect the CO
bandheads from the secondary star in absorption. The CO lines detected in
A0620-00 are observed to be anomalously weak, hinting at possible CNO
processing in the secondary star \citep{b20}. Observations of the XRNe
XTE J1118+480 in outburst reveal a similar situation, with UV spectra
displaying evidence for anomalous abundances of C, N \& O. This points
to CNO processing having also taken place in this system \citep{b119}.  
As explained above, if the C abundance in the secondary star in GS
2000+25 was anomalous as is the case in A0620-00 ([C/H] = -1.5), this
would provide an explanation for their non-detection.

We also note the K-band spectroscopy of WZ Sge \citep{b105} in which
H$_2$ and CO-bandhead emission was detected. These lines are thought to
be produced in dense cool regions as one might expect in the outer
regions of the accretion disc in XRNe \citep{b7}. Hence, it is possible
that our non-detection of the secondary star CO-bandheads in absorption
is due to a combination of an anomalous carbon abundance and infilling
of the CO-bandheads by similar emission lines emanating from the outer
regions of the accretion disc. Unfortunately definitive conclusions
require higher S/N observations as demonstrated in case of A0620-00 by
\citet{b20}.

\subsection{Non-Thermal Jet Emission}\label{jet-discuss}
In an effort to constrain any possible jet contribution to the observed
SEDs, a jet component of the form $\rm f_{\nu} \propto
\nu^{\alpha}$ \symbolfootnote[2]{where the flux is measured in units of
  milli-Janskys: 1mJy $\equiv$ 10$^{-26}$ erg s$^{-1}$ cm$^{-2}$
  Hz$^{-1}$} was added to the model in \ref{results}. The jet is assumed
to be flat and optically thick ($\alpha = 0$), extending from radio
wavelengths and breaking to the optically thin regime ($\alpha < 0$) in
the optical/IR wavelength range (i.e. \citealt{b106}).

There does not appear to be evidence for a significant non-thermal
contribution at NIR wavelengths in any of the systems in our sample. In
A0620-00 and GS2023+338, the jet is seen to contribute at most a few
percent of the observed flux.  Even given the uncertainty in the
\textit{Spitzer} detections at 24$\mu m$ \citep{b6,b82}, a jet will
contribute $\leq$ 5\% and $\leq$ 10\% of the K-band flux in A0620-00 and
GS 2023+338 respectively. It appears that jet emission in quiescence
only becomes an appreciable percentage of the emitted flux as we proceed
to the mid-IR region of the spectrum as noted by \citet{b6}.

\subsection{MIR emission}\label{mir-discuss}
As we proceed to longer wavelengths, the observed flux will become
dominated by emission from the accretion disc/circumbinary disc with a possible
contribution also from a jet, in agreement with the analysis of
\citet{b6}. In the case of the sources with mid-IR spectral
coverage (A0620-00, XTE J1118+480, GS 2023+338, Cen X-4), the
\textit{Spitzer} data favours the presence of a circumbinary disc
contribution to the observed flux. Specifically, whereas a circumbinary disc
is capable of accounting for all the mid-IR flux, a flat jet alone cannot.

\subsection{Black Hole vs Neutron Star XRNe}
Observations of quiescent XRNe with the \textit{Chandra} X-ray
observatory have discovered that the X-ray luminosity of systems
containing black hole primaries appear to be systematically fainter than
those containing neutron stars (see \citealt{b91,b115} and references
therein). This is thought to be due to the fact that the accreting
matter strikes the hard surface in the case of the neutron star whereas
if the primary is a black hole, the matter is simply advected across the
event horizon (although an alternate interpretation envisions the
matter being expelled in a jet/outflow, \citealt{b89}). Hence, it is
worth asking if there is any appreciable difference in the SEDs of the
sources in our sample. 

Even though we only model a single system containing a neutron star
primary, there does not appear to be any discernable difference between
the spectral energy distribution of Cen X-4 and the 6 systems containing
a black hole primary. This is what one would expect if the emission in
the optical/NIR region is dominated by the secondary star and the cooler
outer regions of the accretion disc instead of radiation from processes
taking place closer to the compact object.

\subsection{Black Hole Mass Estimates}
In light of the fact that the NIR flux from XRNe appears to contain a
significant non-stellar contribution, it is worth reconsidering previous
black hole mass estimates made via IR observations, which neglected to
include this non-stellar flux in the analysis. We can estimate the
effect of this additional component using the Eclipsing Light Curve code
(ELC, \citealt{b90}) in combination with the previously determined
system parameters. We then add a contribution from an accretion disc
with a temperature profile of the form T(r) $\propto$ r$^{-0.5}$ and the
inner and outer disc radii set so as to agree with the values determined
in \S \ref{results}. We estimate that for an accretion disc contribution
of only 20\%, the black hole masses measured via NIR photometry have
generally been overestimated by 1 -- 2 M$_{\sun}$ in each case; this
increases to as much as 4 M$_{\sun}$ in the case of a 50\% disc contribution.


\section{Conclusions}
We have shown that there is a significant amount of contamination
present from the accretion disc/circumbinary disc in the IR portion of
the SED of quiescent XRNe. Fits to the SEDs reveal NIR excesses in each
of the 7 systems studied. 
We have also presented new K-band spectroscopy of GS 2000+25, which also
shows some evidence for contamination: this joins two other black hole
XRNe with confirmed IR excesses from spectroscopy (A0620-00 --
\citealt{b20} and  GRO J0422+32 -- \citealt{b21}).

Based on these results, we believe that the currently accepted paradigm,
in which the ellipsoidal variations at NIR wavelengths are assumed to be
undiluted by other sources of flux in the binary, is not valid. We
conclude that the NIR offers no significant advantage over optical
observations in the measurement of ellipsoidal variability (and the
determination of mass ratio and orbital inclination). Indeed, assuming
the contrary introduces large, systematic errors in the mass estimates
for the compact objects in these binary systems.

\bigskip

\noindent \textbf{Acknowledgements}\\
Some of the data presented in this paper was obtained at the Gemini
Observatory, which is operated by the Association of Universities for
Research in Astronomy, Inc., under a cooperative agreement with the
NSF on behalf of the Gemini partnership: the National Science
Foundation (United States), the Science and Technology Facilities
Council (United Kingdom), the National Research Council (Canada),
CONICYT (Chile), the Australian Research Council (Australia), CNPq
(Brazil) and CONICET (Argentina). 

This research made extensive use of the \textit{SIMBAD} database,
operated at CDS, Strasbourg, France and NASA's Astrophysics Data
System. We thank Jerome Orosz for kindly providing us with the ELC code. 
M.T.R. \& P.J.C. acknowledge financial support from Science
Foundation Ireland. 


\vspace{1cm}
\footnotesize{This paper was typeset using a \LaTeX\ file prepared by the 
author}



\begin{thebibliography}{}   

\bibitem[\protect\citeauthoryear{Callanan \& Charles}{1991}]{b11}
  Callanan P.J., Charles P., 1991, MNRAS, 249, 573

\bibitem[\protect\citeauthoryear{Callanan et al.}{1996}]{b12} Callanan
  P.J., Garcia M.R., Filippenko A.V., McLean I., Teplitz H., 1996, ApJ,
  470, 57

\bibitem[\protect\citeauthoryear{Casares et al.}{1993}]{b79} Casares
  J., Charles P.A., Naylor T., Ravlenko E.P., 1993, MNRAS, 265, 834 

\bibitem[\protect\citeauthoryear{Casares et al.}{1997}]{b78} Casares J.,
  Martin E.L., Charles P.A., Molaro P., Rebolo R.J., 1997, NewA, 299,
  310  

\bibitem[\protect\citeauthoryear{Casares}{2005}]{b114} Casares J., 2005,
  in Del Toro Iniesta J.C., Alfaro E.J., Gorgas J.G., Salvador-Sole E.,
  Butcher H., eds, The Many Scales in the Universe: JENAM 2004
  Astrophysics Reviews, Springer, Dordrecht (astro-ph/0503071)  

\bibitem[\protect\citeauthoryear{Charles \& Coe}{2006}]{b74}
  Charles, P.A., Coe M.J., 2006, in Lewin W.H.G., van der Klis
  M., eds, Compact Stellar X-Ray Sources, Cambridge University Press,
  Cambridge (astro-ph/0308020)

\bibitem[\protect\citeauthoryear{Chiang \& Goldreich}{1997}]{b14} Chiang
  E.I., Goldreich P., 1997, ApJ, 490, 368

\bibitem[\protect\citeauthoryear{D'Alessio et al.}{1998}]{b101}
  D'Alessio P., Canto J., Calvet N., Lizano S., 1998, ApJ, 500, 411

\bibitem[\protect\citeauthoryear{D'Alessio et al.}{1999}]{b102}
  D'Alessio P., Calvet N., Hartmann L., Lizano S., Canto J., 1999, ApJ,
  527, 893 

\bibitem[\protect\citeauthoryear{D'Avanzo et al.}{2005}]{b65} D'Avanzo
  P., Campana S., Casares J., Israel G.L., Covino S., Charles P.A.,
  Stella L., 2005, A\&A, 444, 905 

\bibitem[\protect\citeauthoryear{Fender, Gallo \& Jonker}{Fender et
    al.}{2003}]{b89} Fender R.P., Gallo E. \& Jonker J.G., 2003,
    MNRAS, 343, 99

\bibitem[\protect\citeauthoryear{Fender}{2006}]{b106} Fender R., 2006, in
  Lewin W.H.G., van der Klis M., eds, Compact Stellar X-Ray Sources,
  Cambridge University Press, Cambridge (astro-ph/03033339)

\bibitem[\protect\citeauthoryear{Froning \& Robinson}{2001}]{b36} Froning
  C.S., Robinson E.L., 2001, ApJ, 121, 2212

\bibitem[\protect\citeauthoryear{Froning et al.}{2007}]{b20} Froning
  C.S., Robinson E.L., Bitner M.A., 2007, ApJ, 663, 1215

\bibitem[\protect\citeauthoryear{Gallo et al.}{2005}]{b99} Gallo E.,
  Fender R.P., Hynes R.I., 2005, MNRAS, 356, 1017

\bibitem[\protect\citeauthoryear{Gallo et al.}{2006}]{b10} Gallo E.,
  Fender R.P., Miller-Jones J.C.A., Merloni, A., Jonker P.G., Heinz S.,
  Maccarone T.J., van der Klis M., 2006, MNRAS, 370, 1351

\bibitem[\protect\citeauthoryear{Gallo et al.}{2007}]{b82} Gallo E.,
  Migliari S., Markoff S., Tomsick J.A., Bailyn C.D., Berta S., Fender
  R.P., Miller-Jones J.C.A., 2007, arXiv:0707.0028 [astro-ph]

\bibitem[\protect\citeauthoryear{Gelino et al.}{2001a}]{b13} Gelino D.M.,
  Harrison T.E., McNamara B.J., 2001a, AJ, 122, 971

\bibitem[\protect\citeauthoryear{Gelino et al.}{2001b}]{b9} Gelino D.M.,
  Harrison T.E., Orosz J.A., 2001b, AJ, 122, 2668

\bibitem[\protect\citeauthoryear{Gelino \& Harrison}{Gelino et
    al.}{2003}]{b3} Gelino D.M., Harrison T.E., 2003, ApJ, 599, 1254

\bibitem[\protect\citeauthoryear{Gelino et al.}{2006}]{b5} Gelino
  D.M., Balman S., et al., 2006, ApJ, 642, 438

\bibitem[\protect\citeauthoryear{Harlaftis et al.}{1999}]{b80} Harlaftis
  E.T., Collier S., Horne K., Filippenko A.V., 1999, A\&A, 341, 491

\bibitem[\protect\citeauthoryear{Harrison et al.}{2007}]{b104}
  Harrison T.E., Howell S.B., Szkody P., Cordova F.A., 2007, AJ, 133, 162

\bibitem[\protect\citeauthoryear{Haswell et al.}{1993}]{b98} Haswell
  C.A.,Robinson E.L., Horne K., Stiening R.F., Abbott T.M.C., 1993, ApJ,
  411, 802 

\bibitem[\protect\citeauthoryear{Haswell et al.}{2002 }]{b119} Haswell
  C.A., Hynes R.I., King A.R., Schenker K., 2002, MNRAS, 332, 928 

\bibitem[\protect\citeauthoryear{Hauschildt, Allard \& Baron}{Hauschildt
    et al. }{1999a}]{b17} Hauschildt, P.H., Allard F., Baron E., 1999,
    ApJ, 512, 377

\bibitem[\protect\citeauthoryear{Hauschildt et al.}{1999b}]{b16}
  Hauschildt P.H., Allard F., Ferguson J., Baron E., Alexander D., 1999,
  ApJ, 525, 871  

\bibitem[\protect\citeauthoryear{Hodapp et al.}{2003}]{b75} Hodapp K.W.,
  Jensen J.B., Irwin E.M., Yamada H., Chung R., Fletcher K., Robertson
  L., Hora J.L., Simons D.A., Mays W., Nolan R., Bec M., Merrill M.,
  Fowler A.M., 2003, PASP, 115, 1388

\bibitem[\protect\citeauthoryear{Howell et al.}{2004}]{b105} Howell
  S.B., Harrison T.E., Szkody P.,  2004, ApJ, 602, 49

\bibitem[\protect\citeauthoryear{Hynes et al.}{2003}]{b117} Hynes R.I.,
  Charles P.A., Casares J., Haswell C.A., Zurita C., Shahbaz T., 2003,
  MNRAS, 340, 447 
 
\bibitem[\protect\citeauthoryear{Hynes et al.}{2005}]{b7} Hynes R.I.,
  Robinson E.L., Bitner M., 2005, ApJ, 630, 405

\bibitem[\protect\citeauthoryear{Ioannou et al.}{2004}]{b47} Ioannou Z.,
  Robinson E.L., Welsh W.F., Haswell C.A., 2004, ApJ, 127, 481

\bibitem[\protect\citeauthoryear{Kalogera \& Baym}{1996}]{b110}
  Kalogera V., Baym G., 1996, ApJ, 470, 61

\bibitem[\protect\citeauthoryear{Kong et al.}{2002}]{b91} Kong A.K.,
  McClintock J.E., Garcia M.R., Murray S.S., Barret D., 2002, ApJ, 570,
  277 

\bibitem[\protect\citeauthoryear{Lasota}{2007}]{b115} Lasota J.P.,
  2007, CRPhy, 8, 45

\bibitem[\protect\citeauthoryear{Marsh et al.}{1994}]{b86} Marsh T.R.,
  Robinson E.L., Wood J.H., 1994, MNRAS, 266, 137

\bibitem[\protect\citeauthoryear{Menou}{2002}]{b103} Menou K., 2002, in
  Gaensicke B.T., Beuermann K., Reinsch K., eds, The Physics of
  Cataclysmic Variables and Related Objects, ASPC, 261, 387, San
  Francisco 

\bibitem[\protect\citeauthoryear{McClintock et al.}{1995}]{b87}
  McClintock J.E., Horne K., Remillard R.A., 1995, ApJ, 442, 358

\bibitem[\protect\citeauthoryear{McClintock et al.}{2003}]{b88}
  McClintock J.E., Narayan R., Garcia M.R., Orosz J.A., Remillard
  R.A., Murray S.S., 2003, ApJ, 593, 435

\bibitem[\protect\citeauthoryear{McClintock \& Remillard}{2006}]{b2}
  McClintock J.E., Remillard R.A., 2006, in Lewin W.H.G., van der Klis
  M., eds, Compact Stellar X-Ray Sources, Cambridge University Press,
  Cambridge (astro-ph/0306213) 
 
\bibitem[\protect\citeauthoryear{Mitsuda et al.}{1984}]{b15} Mitsuda K.,
  Inoue H., Koyama K., Makishma K., Matsuoka M., Ogawara Y., Shibazaki
  N., Suzuki K., Tanaka Y., 1984, PASJ, 36, 741 

\bibitem[\protect\citeauthoryear{Muno et al.}{2006}]{b6} Muno M.P.,
  Mauerhan J., 2006, ApJ, 648, 135

\bibitem[\protect\citeauthoryear{Narayan}{2002}]{b109} Narayan R.,
  Garcia M., McClintock J., 2002, in Gurzadyan V., Jantzen R., Ruffini
  R., eds, Proc. IX Marcel Grossmann Meeting, Singapore: World Scientific

\bibitem[\protect\citeauthoryear{Narayan}{2008}]{b120} Narayan R.,
  McClintock J.E., 2008, in Abramowicz M.A., Straub O., eds,
  "Jean-Pierre Lasota, X-ray binaries, accretion disks and compact
  stars" New Astronomy Reviews, arXiv:0803.0322 [astro-ph] 

\bibitem[\protect\citeauthoryear{Nielsen}{2007}]{b113} Neilsen J.,
  Steeghs D., Vrtilek S.D., 2007, arXiv:0710.3202 [astro-ph]

\bibitem[\protect\citeauthoryear{Orosz \& Hauschildt}{Orosz et
    al.}{2000}]{b90} Orosz J.A., Hauschildt P.H., 2000, A\&A, 364, 265.

\bibitem[\protect\citeauthoryear{Reynolds et al.}{2007}]{b21} Reynolds
  M.T., Callanan P.J., Filippenko A.V., 2007, MNRAS, 374, 657 

\bibitem[\protect\citeauthoryear{Shahbaz et al.}{1993}]{b66} Shahbaz
  T., Naylor T., Charles P.A., 1993, MNRAS, 265, 655

\bibitem[\protect\citeauthoryear{Shahbaz et al.}{1994}]{b58} Shahbaz
  T., Ringwald F.A., Bunn J.C., Naylor T., Charles P.A., Casares J.,
  1994, MNRAS, 271, 10

\bibitem[\protect\citeauthoryear{Shahbaz et al.}{1996}]{b18} Shahbaz
  T., Bandyopadhyay R., Charles P.A., Naylor T., 1996, MNRAS, 282, 977

\bibitem[\protect\citeauthoryear{Shahbaz et al.}{1999}]{b19} Shahbaz
  T., Bandyopadhyay R., Charles P.A., 1999, A\&A, 346, 82

\bibitem[\protect\citeauthoryear{Shahbaz et al.}{2004}]{b77} Shahbaz T.,
  Hynes R.I., Charles P.A., Zurita C., Casares J., Haswell C.A.,
  Araujo-Betancor S., Powell C., 2004, MNRAS, 354, 31

\bibitem[\protect\citeauthoryear{Torres et al.}{2002}]{b81} Torres
  M.A.P.,  Casares J., Martinez-Pais I.G., Charles P.A., 2002, MNRAS,
  334, 233

\bibitem[\protect\citeauthoryear{Torres et al.}{2004}]{b41} Torres
  M.A.P., Callanan P.J., Garcia M.R., Zhao P., Laycock S., Kong A.K.H.,
  2004, ApJ, 612, 1026

\bibitem[\protect\citeauthoryear{Vacca et al.}{2003}]{b76} Vacca W.D.,
  Cushing M.C., Rayner J.T., 2003, PASP, 115, 389

\bibitem[\protect\citeauthoryear{Wood et al.}{1989}]{b108} Wood J.H.,
  Horne K., Berriman G., Wade R.A., 1989, ApJ, 341, 974

\bibitem[\protect\citeauthoryear{Wood et al.}{1992}]{b107} Wood J.H.,
  Horne K., Vennes S., 1992, ApJ, 385, 294

\bibitem[\protect\citeauthoryear{Yuan \& Cui}{2005}]{c96} Yuan F.,
 Cui W., 2005, ApJ, 629, 408

\bibitem[\protect\citeauthoryear{Zurita et al.}{2003}]{b116} Zurita C.,
  Casares J., Shahbaz T., 2003, ApJ, 582, 369

\end{thebibliography}
\end{document}